\documentclass[aip,pof,reprint,onecolumn]{revtex4-1}

\usepackage{graphicx}
\usepackage[usenames,dvipsnames]{color}

\expandafter\let\csname equation*\endcsname\relax
\expandafter\let\csname endequation*\endcsname\relax
\usepackage{amsmath}
\usepackage{amssymb}

\usepackage{textcomp}
\usepackage{hyperref}

\newcommand{\vk}{von~K\'arm\'an~}

\begin{document}

%\title{How large spheres move in a non-homogeneous and non-isotropic turbulent flow}
%\title{How size and density ratio affect the dynamics of large particles in inhomogeneous and anisotropic turbulence}
%\title{Large inertial particles characteristic time scales in inhomogeneous and anisotropic turbulence}
%\title{How large inertial particles respond to turbulent fluctuations}
\title{Lagrangian velocity and acceleration correlations of large inertial particles in a closed turbulent flow}

%\author{{Nathana\"el Machicoane, Jean-Fran\c{c}ois Pinton, and Romain Volk}\footnote{Author to whom correspondence should be addressed.}}

\author{Nathana\"el Machicoane, and Romain Volk\footnote{romain.volk@ens-lyon.fr}}

\address{Laboratoire de Physique de l'\'Ecole Normale Sup\'erieure de Lyon, CNRS UMR 5672, 46 All\'ee d'Italie, F-69007 Lyon, France}

\begin{abstract}
We investigate the response of large inertial particle to turbulent fluctuations in a inhomogeneous and anisotropic flow. We conduct a Lagrangian study using particles both heavier and lighter than the surrounding fluid, and whose  diameters are comparable to the flow integral scale. Both velocity and acceleration correlation functions are analyzed to compute the Lagrangian integral time and the acceleration time scale of such particles. The knowledge of how size and density affect these time scales is crucial in understanding partical dynamics and may permit stochastic process modelization using two-time models (for instance Sawford's). As particles are tracked over long times in the quasi totality of a closed flow, the mean flow influences their behaviour and also biases the velocity time statistics, in particular the velocity correlation functions. By using a method that allows for the computation of turbulent velocity trajectories, we can obtain unbiased Lagrangian integral time. This is particularly useful in accessing the scale separation for such particles and to comparing it to the case of fluid particles in a similar configuration. 
\end{abstract}

%We conduct a Lagrangian study with particles with diameters comparable to the flow integral scale and slightly heavier or lighter than the fluid.
%\ADD{may allow} to model them as stochastic processes
\maketitle

%%%%%%%%%%%%%%%%%%%%%%%%%%%%%%%%%%%%%%%%%%%%%%%%%%%%%%%%%%%
%%%%%%%%%%%%%%%%%%%%%%%%%%%%%%%%%%%%%%%%%%%%%%%%%%%%%%%%%%%

\section{Introduction}\label{sec:intro}

The dynamics of particles freely suspended in a turbulent flow is a multi scale problem with a wide range of applications such as meteorology, oceanography, and engineering. Among these domains, particle dispersion has been widely studied and still remains an open question concerning large ``material particles'' (i. e. of diameter larger than a portion of flow scales), with a density that can be different from that of the fluid. This stands naturally as a Lagrangian problem where (time) correlation functions are an appropriate tool, being easily accessed and offering direct interpretation. When measuring velocity or acceleration correlation functions over particle trajectories, one can estimate the particle characteristic times, which are typically used in stochastic modelization to achieve prediction of particle dynamics. While only one characteristic time can be enough to simulate the dynamics of an inertial point-like particle immersed in a known flow (in the framework of \cite{Maxey:pof1983,Gatignol:1983}), modeling material particles is often more demanding. A limiting cost can be achieved in direct numerical simulation by use of a corrected Faxen model \cite{Calzavarini2012}, but the fastest way lies in stochastical two-time models in the spirit of Sawford's \cite{sawford:pof1991}. These models make it pertinent to study both the velocity and acceleration time scales of such particles. 

%which are typically used to stochastically model the particle dynamics to achieved some predictions.

In homogenous and isotropic turbulence, the velocity correlations of fluid particles were found to be nearly exponential for time lags larger than the dissipative time of the turbulent flow \cite{snyder1971,sato1987,Yeung1989}. Some studies have also focused on Lagrangian velocity power spectra as such analysis may provide insight into inertial particle response to velocity fluctuations \cite{csanady1963,Obligado2013}. Since the pioneering works using large weather balloons released in the atmosphere \cite{Hanna1980}, or ocean floats \cite{Lien:jfm1998}, modern and controlled experiments have confirmed the presence of an inertial regime in Lagrangian velocity spectra \cite{mordant:njp2004}, and found flow anisotropy existing at the very smallest scales \cite{Ouellette:njp2006}. 

These experiments and simulations consider tracer particles in unconfined and homogeneous turbulence in the absence of any significant mean flow, thus ignoring the influence of the flow topology on particle dynamics. On top of a possible impact of a mean flow on particle dynamics, its presence biases Lagrangian correlation functions and disentangling the signature of turbulence from the mean flow statistics is an important problem in turbulence research, in the general context of close flows.\\ 

The purpose of the present work is to address the dynamics of very large particles freely advected in a confined turbulent flow presenting a large scale structure. These particles, known to become less and less sensitive to the flow fluctuations as compared to the mean flow \cite{Qureshi:prl2007,Brown:prl2009}, preferentially sample the low pressure regions of the flow when their diameter approaches the flow integral scale \cite{machicoane:njp2014}. Complementing these previous works, we investigate Lagrangian velocity and acceleration correlations while tracking the particles for times much larger than the large eddy turnover time of the flow. Our motivation stems from the possibility that particle dynamics may be influenced by the flow topology from the very lowest frequencies down to the frequency cutoff due to the particle size. We investigate the effect of particle size and density on their dynamics, by extracting the particle response times to turbulent fluctuations.

The experiment is performed in a counter rotating \vk flow (briefly described in section \ref{sec:setup}), where particles are tracked optically in the main part of the flow volume. We discuss in section \ref{corrv} the possibility to i) define a turbulent velocity correlation function by removing the mean flow contribution, ii) estimate a Lagrangian integral time. The very high frequency range of the dynamics (i. e. acceleration) and its cutoff due to particle size is then investigated in section \ref{highfreq}. Finally, the scale separation, which is given by the ratio of the velocity and acceleration time scales, for such large particles is discussed in section \ref{discussion}, before a final summary in section \ref{concl}.

\section{Experimental setup}\label{sec:setup}

\begin{figure}[t]
   \centering
   \includegraphics[width=\columnwidth]{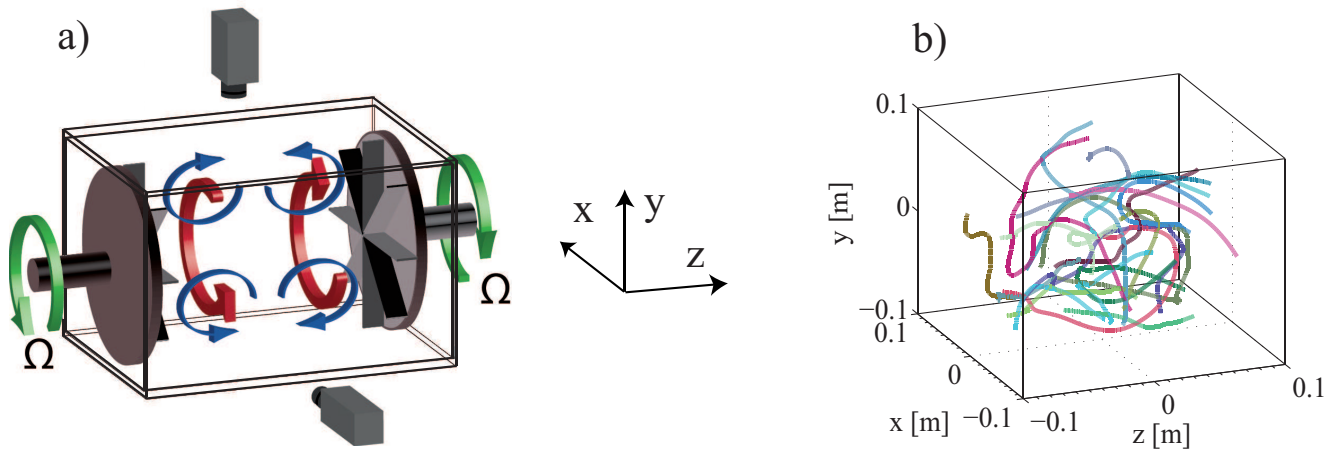}
   \caption{a)~Flow vessel and camera arrangement for 3d particle tracking. Arrows indicate the mean flow geometry with counter rotation in the azimutal direction and meridian recirculation. b)~Sample of $20$ trajectories of $18$ mm polyamide particles obtained with a rotation frequency $\Omega=4$ Hz.}
   \label{fig:KLAC}
\end{figure}

We investigate the dynamics of large inertial particles in a water-filled \vk{} flow, using the same setup as detailed in \cite{machicoane:njp2014,Zimmermann:prl2011}. We chose this system as it presents a well known mean flow with a magnitude of the same order of the turbulent fluctuations. We also note that it shares similarities with industrial mixers and that several previous Lagrangian studies of particles dynamics have been conducted in the center of such vessel \cite{mordant:njp2004,bib:calzavarini2009_JFM,Volk:jfm2011}. The flow is produced by means of two counter rotating disks of radius $R=9.5$~cm located at $z=\pm10$ cm which generate an intense turbulence in a cubic domain of length $H=20\text{ cm}\simeq 2R$. As depicted in Figure \ref{fig:KLAC}(a), the resulting flow has a mean spatial structure composed of a strong azimutal motion near the disks plus a meridian recirculation flow whose origin lies in the pumping motion imposed by the disks. This setup produces non homogeneous and non isotropic turbulence \cite{Ouellette:njp2006, Monchaux:prl2006}, with flow fluctuations stronger near the mid plane ($z=0$). 

The particles are polyamide and polypropylene spheres with diameters $D=[6,\,10,\,18,\,24]$~mm (accuracy $0.01$~mm, Marteau \& Lemari\'e, France) and densities $\rho_\text{PA} = 1.14$~g.cm$^{-3}$ and $\rho_\text{PP} = 0.9$~g.cm$^{-3}$ respectively. All experiments are performed with pure water (kinematic viscosity $\nu= 10^{-6}$~m$^2$.s$^{-1}$) maintained at constant temperature \mbox{$\Theta=20$~\textdegree{}C}. The particles are inertial, and have diameters of the order of the integral length scale $L$ of the flow, much larger than the Kolmogorov length scale (see Table~\ref{table:watergly} for the flow parameters obtained with tracer particles).

Particles are tracked in a very large volume $20\times 20\times 15$~cm$^3$ using 2 high-speed video cameras (Phantom V.12, Vision Research, 1Mpix@6kHz) placed at $90^\circ$ as shown in Figure \ref{fig:KLAC}(a). Both cameras observe the measurement volume with a resolution $800~\text{x}~768$ pixels, limiting the maximum length of trajectories to $13900$ frames. We performed experiment with a moderate sampling frequency adjusted in the range $f_\text{sampling} \in [1600, ~3000]$ Hz (depending on the disks velocity) to obtain trajectories with mean duration $\langle \Delta t \rangle \simeq 0.5/ \Omega$ (Figure \ref{fig:KLAC}(b)). We can then obtain velocity and acceleration trajectories by convolution of a filtering-differentiating Gaussian kernel (as was done in \cite{machicoane:njp2014}) allowing for the computation of velocity/acceleration correlation functions over meaningful times. \\

\begin{table}[tbh]
\begin{center}
\begin{tabular}{|l|c|c|c|c|c|c|c|c|}
\hline
\textbf{$\Omega$}&\textbf{$u' [m.s^{-1}]$}&\textbf{$\varepsilon_{power} [W.kg^{-1}]$}&\textbf{$\eta [\mu m]$}&\textbf{$\tau_\eta [ms]$}&\textbf{$Re$}&\textbf{$R_\lambda$}&\textbf{$L [cm]$}\\\hline
\textbf{2 Hz}&0.25&0.48&38&1.44&1.2~$10^6$&350&3.2\\\hline
\textbf{3 Hz}&0.38&1.67&28&0.77&2.4~$10^6$&430&3.2\\\hline
\textbf{4 Hz}&0.52&4.03&22&0.5&3.6~$10^6$&520&3.4\\\hline
\end{tabular}
\caption{Experimental parameters. $\Omega$: rotation frequency of the counter rotating disks. $u'=\sqrt{((u'_x)^2+(u'_y)^2+(u'_z)^2)/3}$: fluctuating velocity averaged over components measured from tracer dynamics in a large volume around the geometrical center. $\varepsilon$: energy dissipation estimated from the electrical power consumption of the motors (see \cite{machicoane:njp2014} for more details). $\tau_\eta \equiv (\nu/\varepsilon)^{1/2}$: Kolmogorov time scale. $\eta \equiv (\nu^3/\varepsilon)^{1/4}$: Kolmogorov length scale. $Re\equiv (2 \pi R^2\Omega)/\nu$: Reynolds number computed using the disk tip velocity. $R_\lambda=\sqrt{15{u'}^4/\varepsilon\nu}$: Reynolds number based on the Taylor micro scale. $L={u'}^3/\varepsilon$: the estimated integral length scale.}
\label{table:watergly}
\end{center}
\end{table}

\section{Estimating turbulent velocity  correlations}\label{corrv}

\begin{figure}[htbp]
  \begin{center}
  \includegraphics[width=\columnwidth]{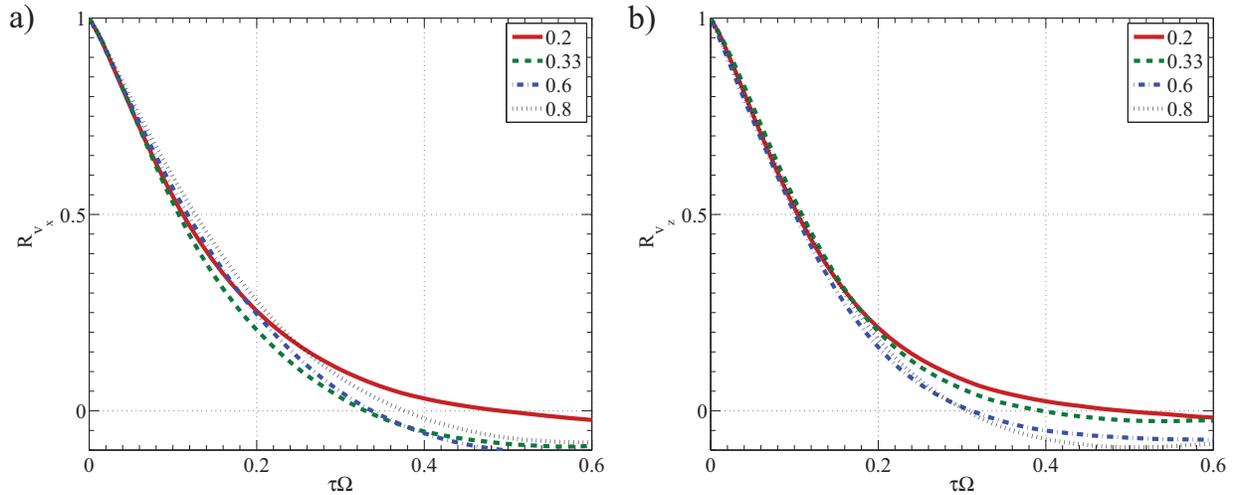}
%  \includegraphics[width=.45\columnwidth]{fig_new/corr_v_total_PA_4Hz_x.pdf}
%    \includegraphics[width=.45\columnwidth]{fig_new/corr_v_total_PA_4Hz_z.pdf}\\
%      a) \hspace{7cm} b)\\
    \caption{a) Normalized auto-correlation functions of the transverse velocity $v_x$, noted $R_{v_x}$, for polyamide particles with diameters $D/L=0.2$ (solid line), $D/L=0.33$ (dashed line), $D/L=0.6$ (dashed-dotted line), $D/L=0.8$ (dotted line), and a rotation frequency $\Omega=4$ Hz. b) Corresponding normalized auto-correlation functions of the axial velocity $v_z$.}
    \label{cvtot}
  \end{center}
\end{figure}

The aim of this section is to compute the characteristic time of particle velocity fluctuations. The next session will consider the response time to high frequency fluctuation, linked to acceleration. Studying how both times change with the particle characteristics yield then an understanding of size and density effect on particle dynamics.
In order to compute velocity correlations, one must track particles for durations longer than the large eddy turnover time of the flow $T=1/\Omega$. Particles will then explore a significant part of the flow during one track and may be lost for small durations when hidden by other particles, cutting the tracks in an ensemble of smaller trajectories. Though this is a severe limitation when computing Lagrangian spectra, this does not introduce any bias in the estimation of the correlation functions providing the particles can be identified before and after being lost. We thus run the experiments with one particle per diameter in the flow volume so that we reduce the number of experiments while being able to identify all particles. Figures \ref{cvtot}(a,b) display the resulting normalized auto-correlation functions, $R_{v_i}(\tau)=\langle v_i(t)v_i(t+\tau) \rangle /\langle v_i^2 \rangle$ ($i=x,z$), for the four sizes of polyamide particles at the largest Reynolds number considered. \\ 

These figures show the particle size has a weak impact on velocity correlation functions for all the velocity components. All curves have a very similar shape at short times, and it is very difficult to differentiate them for particles larger than $10$ mm ($D/L=0.33$) as all auto-correlation functions become negative when $\tau \Omega \sim 0.4$. After crossing zero, the functions then tend towards zero on a longer time scale (with eventual oscillations), which is not shown here as the statistics are less converged for large time lags $\tau$. This zero-crossing behavior is not due to any finite volume bias as discussed for Lagrangian measurements performed in the homogeneous central region of \vk flows \cite{mordant:njp2004}. It is due to the fact that particles are tracked in a very large portion of a bounded flow. Indeed, in this case, the Lagrangian integral time, $T_L= \int_0^\infty R_{v}(\tau) d\tau$, must be zero for a bounded flow \cite{tennekes1972book}. This result means one must disentangle the contributions of the mean flow from the turbulent fluctuations in order to estimate turbulent velocity correlations and a Lagrangian integral time.

\begin{figure}[htbp]
  \begin{center}
  \includegraphics[width=\columnwidth]{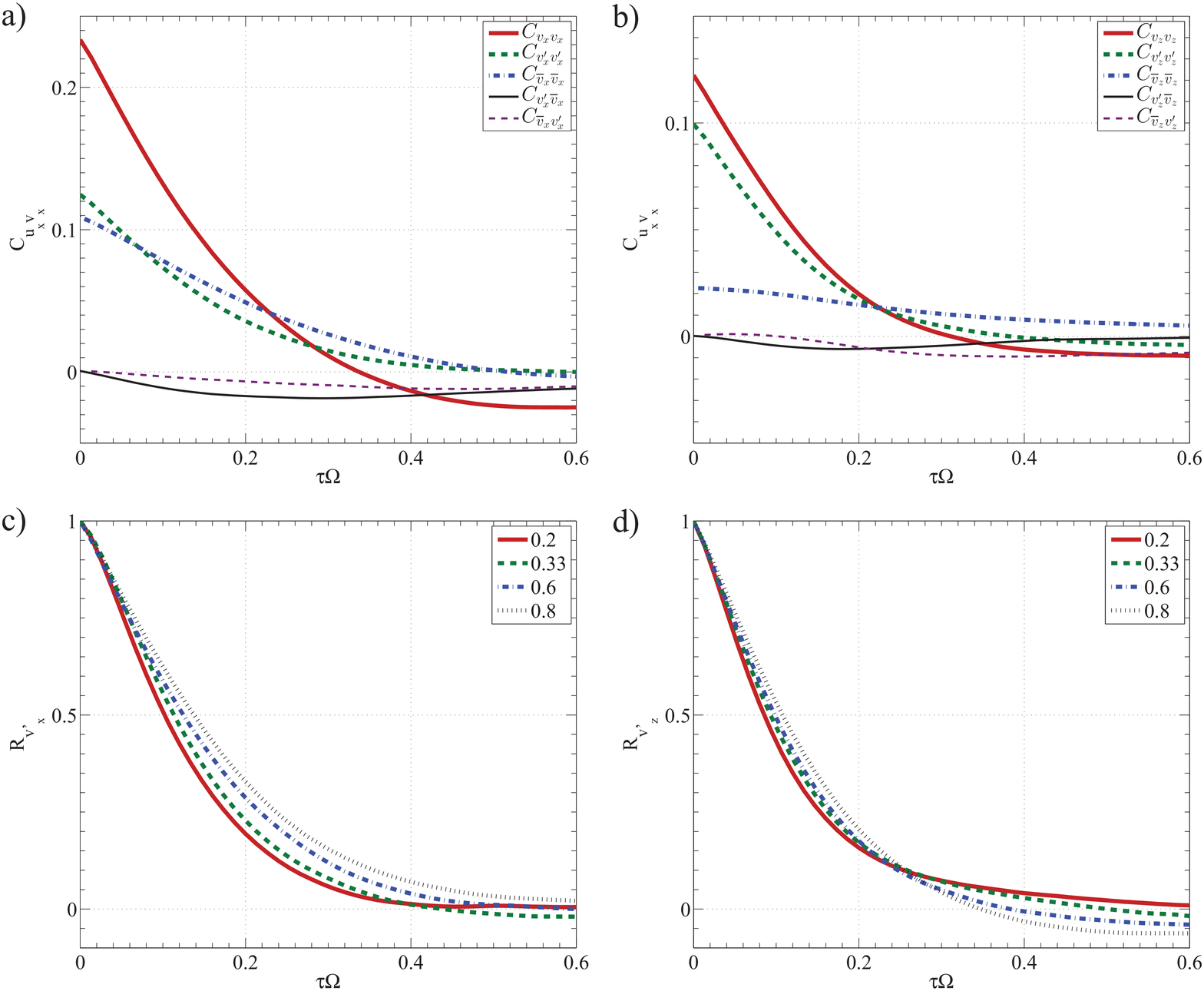}
%  \includegraphics[width=.45\columnwidth]{fig_new/comp_corr_v_PA_6mm_4Hz_x.pdf}
%  \includegraphics[width=.45\columnwidth]{fig_new/comp_corr_v_PA_6mm_4Hz_z.pdf}\\
%    a) \hspace{7cm} b)\\
%  \includegraphics[width=.45\columnwidth]{fig_new/corr_vp_PA_4Hz_x.pdf}
%  \includegraphics[width=.45\columnwidth]{fig_new/corr_vp_PA_4Hz_z.pdf}\\
%    c) \hspace{7cm} d)\\
\caption{a) Correlation functions of the transverse velocity (see equation \ref{eq:cvv}) for polyamide particles with $D/L=0.6$ and a rotation frequency $\Omega=4$ Hz. $C_{v_xv_x}$ (thick solid line), $C_{v'_xv'_x}$ (dashed line), $C_{\overline{v}_x\overline{v}_x}$ (dashed-dotted line), $C_{\overline{v}_xv'_x}$ (dotted line), $C_{v'_x\overline{v}_x}$ (thin solid line). b) Corresponding correlation functions for the axial velocity. c) Normalized auto-correlation functions of the fluctuating velocity $v'_x$, noted $R_{v'_x}$, for polyamide particles with diameters $D/L=0.2$ (solid line), $D/L=0.33$ (dashed line), $D/L=0.6$ (dashed-dotted line), $D/L=0.8$ (dotted line), and a rotation frequency $\Omega=4$ Hz. d) Corresponding normalized auto-correlation functions of the axial fluctuating velocity $v'_z$.}
    \label{corr_v_prime_v_tot_D}
  \end{center}
\end{figure}

To achieve this, we compute the stationary Eulerian flow for each particle type as the ensemble average $\overline{\mathbf{v}}_\text{E}(x,y,z)=\big{(}\overline{v}_x,\overline{v}_y,\overline{v}_z\big{)}$, and define the particle velocity components as the sum of the local mean flow value at the particle location $\mathbf{x}(t)$ and the fluctuating velocity: $v_i(t)=v'_i(t)+\overline{v}_i(\mathbf{x}(t))$. More details about this Eulerian Lagrangian conditioning can be found in \cite{machicoane:njp2014}. The auto-correlation function of $v_i$ is then the sum of four terms :

\begin{equation}
\begin{split}
\langle v_i(t)v_i(t+\tau) \rangle = \langle v'_i(t)v'_i(t+\tau) \rangle +  \langle \overline{v}_i(\mathbf{x}(t))\overline{v}_i(\mathbf{x}(t+\tau)) \rangle \\
+ \langle \overline{v}_i(\mathbf{x}(t))v'_i(t+\tau) \rangle + \langle v'_i(t)\overline{v}_i(\mathbf{x}(t+\tau)) \rangle,
\end{split}
\end{equation}

which can be rewritten as
\begin{equation}\label{eq:cvv}
C_{v_iv_i}=C_{v'_iv'_i}+C_{\overline{v_i}\overline{v_i}}+C_{\overline{v_i} v'_i}+C_{v'_i\overline{v_i}},
\end{equation}

where $C_{u_iv_i}$ stands for the covariance of $u_i$ and $v_i$ (\textit{i.e.} non normalized correlation function). We compute the five covariances for all particle types, and display the results for the transverse and axial velocity components of a polyamide particle with size $D/L=0.6$ in Figures \ref{corr_v_prime_v_tot_D}(a,b) respectively. Whatever the component, we observe that the cross correlations $C_{\overline{v} v'}$ and $C_{v'\overline{v}}$ are much smaller than $C_{\overline{v}\overline{v}}$ and $C_{v'v'}$. This property, valid whatever the particle size and density ratio, implies that fluctuations experienced by particles can be considered to be uncorrelated with their dynamics due to the mean flow even if particles do not sample the flow homogeneously. For the transverse component, we also find that $C_{v'v'}$ and $C_{\overline{v}\overline{v}}$ are of the same order of magnitude. The \textit{particles} dynamics is reflecting a key property of the counter rotating \vk \textit{flow} by which mean velocity and transverse fluctuations are of the same order of magnitude. The result concerning the axial component is more surprising because we observe the total velocity correlation is dominated by the turbulent contribution $C_{v'_zv'_z}$. This result also holds for other particle sizes although less apparent for smaller particles which are not trapped in regions of small mean axial velocity (\textit{i. e.} toroidal regions close to the disks).

As the proposed decomposition is valid whatever the particle size, we display in Figure \ref{corr_v_prime_v_tot_D}(c) the resulting auto-correlation functions of the turbulent transverse velocity, $R_{v'_x}$, for all particle diameters. As opposed to the total velocity, the curves have similar shapes but a clear ordering appears for different particle diameters. The functions time scales indeed increase with particle size, but they no longer cross zero (up to statistical convergence) as if the dynamics were unbounded. The result is different for the axial component where we find that the shape of $R_{v'_z}$ strongly evolves with the particle size as demonstrated in Figure \ref{corr_v_prime_v_tot_D}(d). The functions become more and more curved as the particle size increases and it becomes evident that no rescaling can put theses functions on a single master curve. Indeed, while the functions are quite ordered at small times, the changes of shape lead to a point where all functions cross each other (around $\tau\Omega=0.25$), resulting in a reversal of the order. The functions shape for larger particles yields a zero crossing and eventual oscillations, as was observed before for the total velocity (figure \ref{cvtot}). Through the method seems to work well for the smallest particle considered, it appears not to be sufficient for larger particles for which another bias is present. This component is indeed more sensitive to preferential sampling, which explains that $R_{v'_z}$ oscillates when particles are large enough to experience coming and going motions between trapping regions of small axial mean flow with high gradient in fluctuating velocity.  

\begin{figure}[htbp]
  \begin{center}
  \includegraphics[width=\columnwidth]{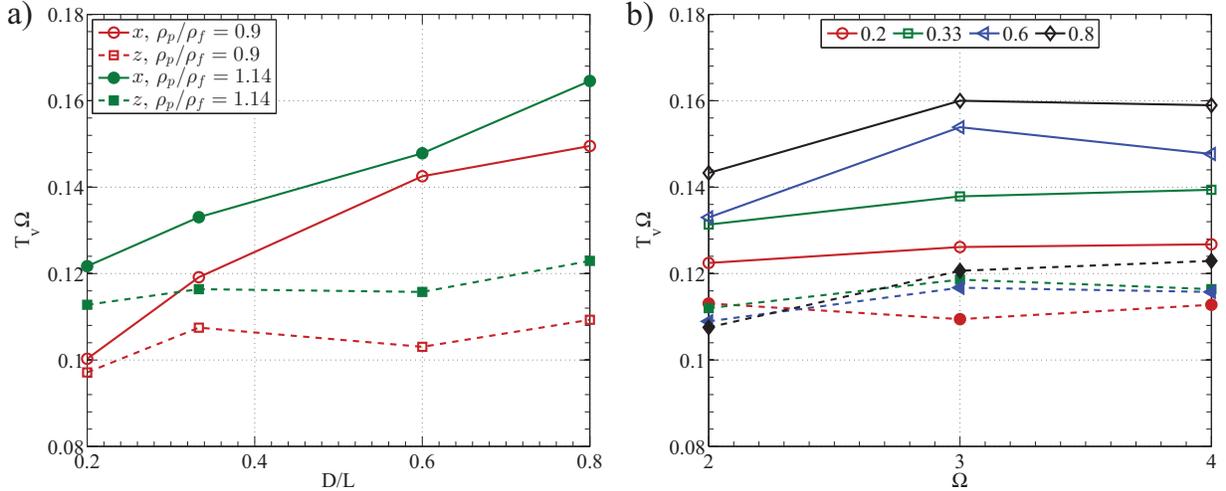}
    \caption{a) Evolution of the velocity integral time as a function of particle size for polyamide particles (filled symbols) with relative density $1.14$ and polypropylene particles (open symbols) with relative density $0.9$. {\large $\circ$} and solid line: transverse  component $x$. $\square$ and dashed line: axial  component $z$. b) Evolution of the velocity integral time as a function of the rotation frequency for polyamide particles along the transverse component $x$ (continuous line, filled symbols) and axial component (dotted line, empty symbols) for different diameters (colours and symbols).}
    \label{Tv}
  \end{center}
\end{figure}

As it is difficult to achieve statistical convergence of velocity correlations for large time lags we define a Lagrangian integral time, $T_{v}=\int_0^{\tau_{95}} R_{v'}(\tau) d\tau$, where $\tau_{95}$ is the time needed for $R_{v'}(\tau)$ to decrease by $95\%$. This ensures one can measure the integral time for all velocity components without accounting for the negative part of $R_{v'}(\tau)$. Figure \ref{Tv} shows the transverse velocity time scale increases by $30\%$ when increasing the particle size from $D/L=0.2$ to $0.8$, which is consistent with other results obtained using a corrected Faxen model to simulate the dynamics of isodense particles with sizes larger than the Kolmogorov scale \cite{Calzavarini2012}. As a result of the velocity correlation's anisotropy, we find $T_{v}$ is strongly anisotropic, being smaller for the axial component as compared to the transverse one. Furthermore, no substantial variation of $T_{v}$ with $D$ is detected for the axial velocity because the shape of $R_{v'_z}$ strongly changes at increasing particle size (which biases the measure of $T_v$). Up to now, only results at the highest Reynolds number have been presented. Indeed, the turbulence is fully developed, meaning that the $Re$ dependence of time scales is fully captured by the rotation frequency $\Omega$. This is easily checked when looking at the evolution of $T_{v}\Omega$ with $\Omega$ (figure \ref{Tv}(b)). We indeed observe no significant variations whatever the component or particle diameter considered.

Concerning the influence of relative density, we find that light particles present a similar behavior to the one of heavy particles but with an integral time $10\%$ smaller, irrespective of the velocity component. In the spirit of Tchen-Hinze theory \cite{hinze}, this may be understood as an effect of the added mass force. Indeed it is in agreement with previous measurements of acceleration magnitude which show the rms value of acceleration to be proportional to $\beta=3\rho_f/(2\rho_p+\rho_f)$, where $\rho_p$ is the particle density \cite{machicoane:njp2014}. In this framework, high frequency fluctuations appear to be amplified or attenuated depending on the value of $\beta$ ($\beta_\text{PP}=1.07$, while $\beta_\text{PA}=0.91$) such that light particles always have smaller correlation times (as will be shown in section \ref{highfreq}). 

%This is indeed visible in Figure \ref{lowspec}(d) which shows light particle position spectra are more energetic than those of heavier particles in the high frequency range. 

%\textbf{si on met une euqation c'est $\frac{Dv}{Dt}=\beta\frac{Du}{Dt}+\frac{1}{\tau_p}(u-v)$ et il faut expliquer que $\tau_p\sim1/\beta$ et donc à même $D$ changer $\beta$ change $Dv/dt$ d'un facteur $\beta$?}
\section{High frequency cutoff}\label{highfreq}

\begin{figure}[h]
  \begin{center}
  \includegraphics[width=\columnwidth]{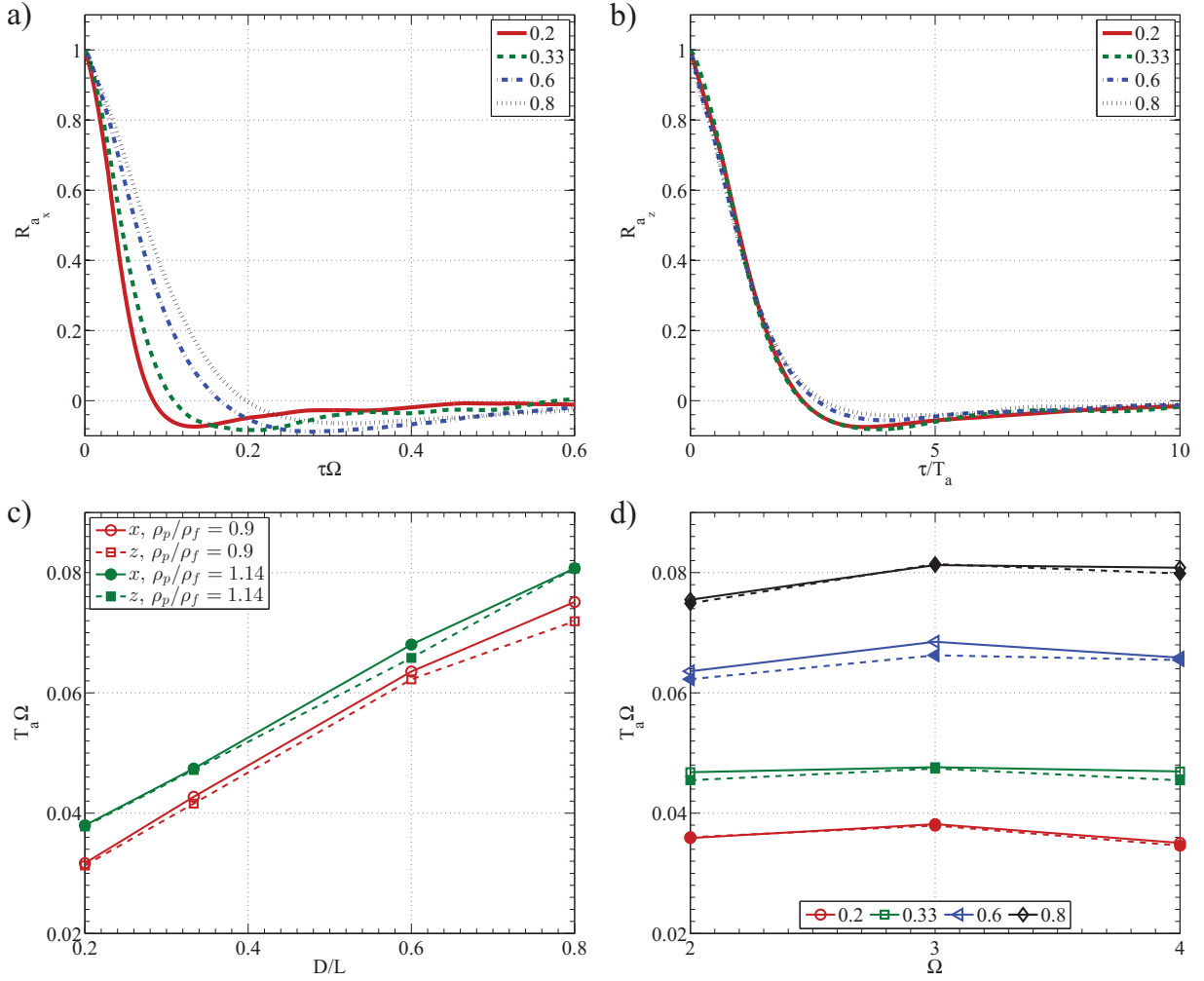}
%      \includegraphics[width=.45\columnwidth]{fig_new/corr_acc_y_3Hz_PA.pdf}
%    \includegraphics[width=.45\columnwidth]{fig_new/corr_acc_norm_z_4Hz_PA.pdf}\\
%      a) \hspace{7cm} b)\\   
%        \includegraphics[width=.45\columnwidth]{fig_new/tau_acc_fct_D.pdf}
%    \includegraphics[width=.45\columnwidth]{fig_new/tau_corr_a_fct_freq.pdf}\\
%      c) \hspace{7cm} d)\\
    \caption{a) Axial acceleration auto-correlation functions of polyamide particles, $R_{a_z}$ for polyamide particles with diameters $D/L=0.2$ (solid line), $D/L=0.33$ (dashed line), $D/L=0.6$ (dashed-dotted line), $D/L=0.8$ (dotted line), and a rotation frequency $\Omega=4$ Hz. b) Same figure as (a) but with the time lag $\tau$ normalized by the acceleration correlation time $T_a=\int_0^{\tau_0} R_{a}(\tau) d\tau$, where $R_a(\tau_0)=0$. c) Evolution of the acceleration correlation time, $T_a$, as a function of particle size for polyamide particles with relative density $1.14$ ({\large $\bullet$} $a_x$, $\blacksquare$ $a_z$)  and polypropylene particles ({\large $\circ$} $a_x$, $\square$ $a_z$) with relative density $0.9$. d) $T_a\Omega$ as a function of $\Omega$ for polyamide particles for the transverse component $x$ (continuous line, filled symbols) and axial component (dotted line, empty symbols) for different diameters (colours and symbols).}
    \label{highspec}
  \end{center}
\end{figure}

In order to estimate the ``dissipative time" of particle dynamics, we now focus on small time scales and investigate acceleration correlation functions. Acceleration is a short correlated quantity as compared to velocity and is only weakly affected by the mean flow, hence only results for the total particle acceleration are presented. Figure \ref{highspec}(a) presents the acceleration autocorrelation functions of polyamide particles for the four diameters investigated. One can clearly observe that particle acceleration decorrelates in a similar manner albeit at larger and larger time scales as the particle size increases. Indeed, normalizing the time lag axis $\tau$ by the acceleration correlation time $T_a$ (defined as the integral of the positive part of $R_a$ as introduced for small particles \cite{bib:calzavarini2009_JFM,Volk:jfm2011}), we find that all functions almost collapse on a master curve (figure \ref{highspec}(b)), even when looking at the axial component. %On time lags larger than $\tau \sim 1.5 ~T_a$, one can still differentiate between small ($6$ and $10$ mm) and large ($18$ and $24$ mm) particles which become heavily influenced by particle trapping. However, we observe acceleration correlation functions still have similar shapes whatever the component or the particle size and density. 

%that particle acceleration decorrelates with a larger and larger time scale as the particle size increases, but in a similar manner

Figure \ref{highspec}(c) shows $T_a$ increases at increasing particle size in a more pronounced fashion than $T_v$ ($T_a$ is approximately multiplied by 2.5 when $D$ increases fourfold). This increase of acceleration correlation time is not incompatible with the one found for much smaller material particles \cite{Volk:jfm2011} (where $T_a/\tau_\eta$ increases by 2.8 when $D/\eta$ increases from 10 to 40). Contrary to observations made for the velocity, transverse accelerations have time scales barely larger than axial ones due to the fact that acceleration is more isotropic, as was observed for fluid particles \cite{mordant2004acc}. This quasi-isotropy stands whatever the particle size or density, and, as it is the case for previous results, the Reynolds number dependence is fully captured by normalizing quantities with the rotation frequency $\Omega$ (as seen on figure \ref{highspec}(d)).

\begin{figure}[h]
  \begin{center}
  \includegraphics[width=\columnwidth]{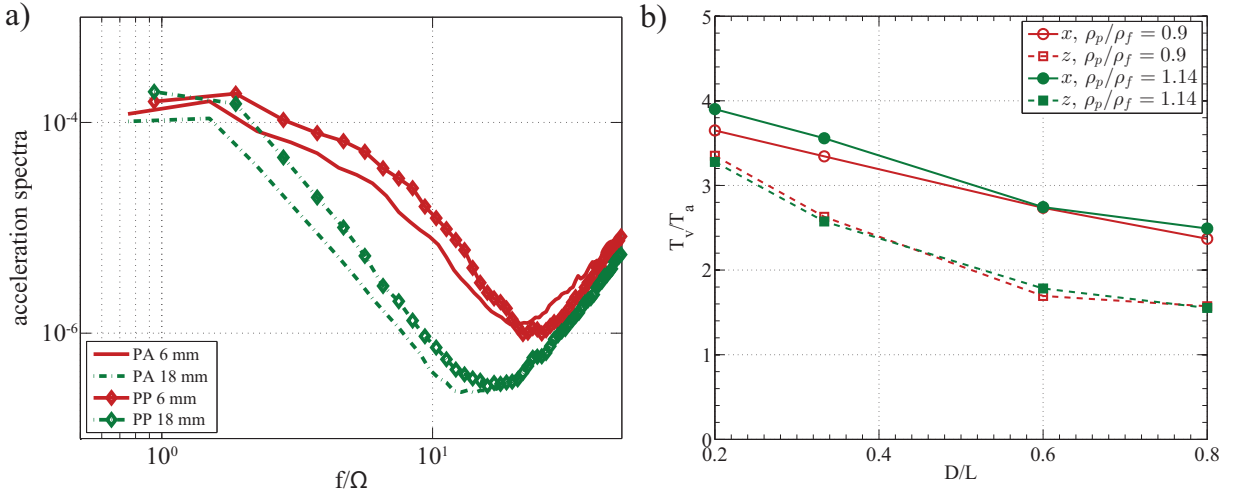}
%      \includegraphics[width=.45\columnwidth]{fig_new/NJP2_figure3b_PAPP_6_18_nath.pdf}
%    \includegraphics[width=.45\columnwidth]{fig_new/ratio_taus_fct_D.pdf}\\
%      a) \hspace{7cm} b)\\   
    \caption{a) Transverse raw acceleration power spectra for polyamide particles with diameters $D/L=0.2$ (solid line), $D/L=0.6$ (dashed line) and polypropylene particles (diamonds) with diameters $D/L=0.2$ (solid line), $D/L=0.6$ (dashed line), with a rotation frequency $\Omega=4$ Hz. b) Evolution of the ratio of the velocity and acceleration correlation time, $T_v/T_a$, as a function of particle size for polyamide particles with relative density $1.14$ ({\large $\bullet$} $a_x$, $\blacksquare$ $a_z$)  and polypropylene particles ({\large $\circ$} $a_x$, $\square$ $a_z$) with relative density $0.9$.}
    \label{highspec2}
  \end{center}
\end{figure}

Varying the density in the range $\rho_p/\rho_f \in [0.9,~1.14]$, one finds light particles have a time scale approximately $10\%$ shorter than heavier particles as was observed for the velocity (figure \ref{highfreq}(d)). A closer analysis of the density ratio effect on particle dynamics can be made by use of the raw acceleration power spectra. Even if they are directly linked to correlation functions (by a Fourier transform), we find them more appropriate to compare light particles dynamics to the one of heavier particles through all time scales. Figure \ref{highspec2}(a) shows indeed that whatever the particle size, acceleration power spectra are more energetic for a given light particle than for its heavy counterpart. Of course this yields higher/lower rms acceleration values for light/heavy particles (by a factor $\beta=3\rho_f/(\rho_f+2\rho_p)$ as shown in \cite{machicoane:njp2014}), which correspond to the integral of the power spectra. 

Whatever the density, spectra exhibit a sharp cutoff in the range $2 \leq f/\Omega \leq 20$ with a cutoff frequency, $f_c$, which decreases at increasing particle size (linked to the increase of $T_a$ with $D$). When a power law is tested, one finds $\phi_{a}(f) \propto f^{-\alpha}$ with an exponent $\alpha \simeq 3$. The present decrease is faster than that observed for ocean floaters \cite{Lien:jfm1998}, or as predicted by Sawford's two-times model for which acceleration spectra decrease as $f^{-2}$ beyond the cutoff frequency \cite{sawford:pof1991}. Accordingly, spectra do not exhibit any plateau compatible with a Kolmogorov scaling as observed for small particles, likely due to a lack of scale separation between the velocity integral time $T_v$ and the acceleration time scale $T_a$. Note that the increase in the raw acceleration power spectra for high frequencies corresponds to noise coming from position measurement.

\section{Discussion}\label{discussion}
%While this confirms that no decomposition is indeed needed to study particle acceleration, this small $T_v/T_a$ ratio explains why sensitivity to the mean flow can still be felt for larger time lags (see the separation between smaller and larger particles in figure \ref{highspec}(b))
Finding acceleration less sensitive to trapping effects or to the mean flow geometry can be expected as it is a small-scale quantity. However, with the small scale separation $T_v/T_a$ found for these large particles, it is surprising that theses results still stand (which is confirmed by the fact that no decomposition is indeed needed to study particle acceleration). Few Lagrangian studies have measured both the velocity and acceleration time scales of fluid particles in turbulent flows and we will focus our comparison with the work of \cite{mordant:njp2004} which took place in a similar setup. Note that the particles considered in \cite{mordant:njp2004} are slightly larger than the Kolmogorov length scale. The velocity autocorrelation functions of such tracers have an exponential shape, with the Lagrangian time $T_L$ as characteristic time scale, yielding $T_v^\text{tracer}\simeq T_L$. The authors indicated that the acceleration correlation functions cross zero at a value approximately 4 times larger than the value $2.2\tau_\eta$ commonly reported \cite{Yeung1989,Volk:jfm2011,mordant2004acc} (an effect of the particles' finite size). Using a better estimate of the zero crossing time as $2 \tau_\eta$, and assuming the positive part of the acceleration function is triangular, we then have $T_a^\text{tracer}\simeq\tau_\eta$. In the range of Taylor micro scale Reynolds number $500<R_\lambda<1000$ explored, we can then estimate from their data $70<T_v^\text{tracer}/T_a^\text{tracer}<130$, increasing linearly with $R_\lambda$ \cite{sawford2013}, which is consistent with the fact that the inertial range (hence the scale separation) grows at increasing Reynolds number. In the case of the large particles considered in the present study, we find ratios between 1.5 and 4 depending on the particle size and acceleration component (figure \ref{highspec2}(b)). This is indeed a much smaller scale separation for such particles, which is enlightening for modeling their dynamics, using for instance a two-time model (\cite{sawford:pof1991} for instance). We note that, as both $T_v$ and $T_a$ scale with $\Omega$ in the same way, there is no Reynolds number dependence of the scale separation for large particles, in opposition to the case of fluid particles. This is in agreement with the fact that the acceleration magnitude of large particles does not follow the classical $a_\text{rms}\sim\varepsilon^{3/4}\nu^{-1/4}$ (see for instance \cite{Brown:prl2009}) but was found to scale with the forcing time scale $a_\text{rms}\sim R\Omega^2$ \cite{machicoane:njp2014}. $D/L$ stands then as the good parameter, as the integral length scale is approximately constant with the Reynolds number, while the Kolmogorov length scale decreases with $Re$. Indeed, plotting $Ta\Omega$ (or just $Ta$) as function of $D/\eta$ would yield different sets of curves, while one master curve emerges when plotted against $D/L$.

%We should also consider than the particles used in this study are slightly larger than the Kolmogorov length scale: their diameter is 10 times $\eta$ at $R_\lambda=570$, which means that the ratio is underestimated (on top of the fact that we overestimate $T_a$). We can however extrapolate the value of $T_L$ for tracers by use of the expression $T_L=2v'/C_0\varepsilon$ where $C_0=7$ is the Lagrangian correlation constant and use the fact that accelration correlation of tracers cross zero at $2.2\tau_\eta$. This yields a scale separation of 30 (instead of 18) at $R_\lambda=570$.

The particle density ratio, found to increase both velocity and acceleration time scales by approximately the same factor, play indeed no role on the $T_v/T_a$ ratio (figure \ref{highspec2}(b)). The observed decrease of this ratio confirms that velocity time scales increase slower than that of acceleration, which is highlighted by the axial component where velocity time scales tend towards a constant value. The anisotropy of the scale separation is therefore attributed to velocity anisotropy, which is inherent to \vk flows. We believe that, as the impact of the mean flow is totally removed for the transverse components, its evolution is more representative. It would be interesting to compare this decrease of the scale separation to particles of size in or below the inertial range. To our knowledge, previous studies of smaller material particles (typically in the range $1<D/\eta<50$ \cite{bib:calzavarini2009_JFM,Volk:jfm2011}), only report measurements of acceleration time scales, as the velocity time scale demands much longer trajectories. %This does not allow us to further comment on the evolution of $T_v/T_a$ with particle size, which could constitute an interesting study.

\section{Summary}\label{concl}
%%%%%%%%%%%%%%%%%%%%%%%%%

Using 3d particle tracking, we have studied the characteristic time scales of very large inertial particles in a non homogeneous fully turbulent flow. Lagrangian tools have shown the strong influence of particle size on their dynamics. Indeed, for these large objects, we found both acceleration and velocity time scales to be strongly increasing with the particle size. This is the signature of a modification of the particle dynamics over the whole range of frequencies, as shown by the Lagrangian power spectra of the particle acceleration. Frequencies higher than the injection frequency displayed strong cutoffs in acceleration spectra due to particle size, corresponding to the observed increase of acceleration time scales $T_a$ with the particle size. Although this is in agreement with previous measurements of finite size particle acceleration \cite{bib:calzavarini2009_JFM,Volk:jfm2011}, acceleration spectra were found to decrease much faster than $f^{-2}$ as reported for ocean floats with much smaller $D/\eta$ ratios \cite{Lien:jfm1998}.\\

In order to separate the effects of size and density, we varied the particle relative density and compared systematically the results obtained for particles $10\%$ lighter than the fluid to the one of particles $14\%$ heavier. For all quantities, the role of density was found much less influential than particle size. This is a specificity of particles much larger than the Kolmogorov length scale, for which density appears only to finely tune the dynamics, whose main features (trapping and response to turbulent fluctuations) are imposed by the particle size. As a result, the observed variation in the statistics can be qualitatively explained through the action of the added mass force.\\ 

Our study has investigated the dynamics of particles over the whole volume of a confined flow and tracked them for durations larger than the large eddy turnover time. As a consequence, velocity correlations were biased by contributions from the mean flow so that the Lagrangian integral time was zero. By removing the local value of the mean flow along each trajectory to estimate fluctuating velocity correlations, we were able to quantify the effects of size and density on the Lagrangian integral time. As particles above a certain size develop coming and going motions in the axial direction in such flows \cite{machicoane:njp2014}, the autocorrelation functions of the axial velocity are more heavily biased. Indeed, the shape of the autocorrelation function of the fluctuating velocity remained strongly influenced by the particle size such that the Lagrangian integral time remained the same up to measurement errors (in the axial direction only). This shows how difficult it is to disentangle the influence of the mean flow and turbulent fluctuations in a non homogeneous turbulent flow.\\

%is more heavily biased
%was still strongly influenced by the particle size so that the Lagrangian integral time remained the same

\noindent{\bf Acknowledgments} The authors want to thank Peter Huck for his reading of the manuscript, and Jean-Fran\c{c}ois Pinton, Mickael Bourgoin, Nicolas Mordant and Javier Burguete for stimulating discussions. This work is supported by French research program ANR-12-BS09-0011 ``TEC2'' and ANR-13-BS09-0009 ``LTIF''.

\section*{References}
%\bibliographystyle{plain}
%\bibliography{biblio_Lsphere}

\providecommand{\newblock}{}
\begin{thebibliography}{}
\expandafter\ifx\csname url\endcsname\relax
  \def\url#1{{\tt #1}}\fi
\expandafter\ifx\csname urlprefix\endcsname\relax\def\urlprefix{URL }\fi
\providecommand{\eprint}[2][]{\url{#2}}

\end{thebibliography}


\begin{thebibliography}{99}


\bibitem{Maxey:pof1983}
Maxey M~R and Riley J~J 1983 {\em Physics of Fluids\/} {\bf 26} 883--889

\bibitem{Gatignol:1983}
Gatignol R 1983 {\em Journal de M\'ecanique Th\'eorique et Appliqu\'ee\/}

\bibitem{Calzavarini2012}
Calzavarini E, Volk R, L\'{e}v\^{e}que E, Pinton J~F and Toschi F 2012 {\em
  Physica D\/} {\bf 241} 237--244

\bibitem{sawford:pof1991}
Sawford B~L 1991 {\em Physics of Fluids A\/} {\bf 3} 1577--1586

\bibitem{snyder1971}
Snyder W~H and Lumley J~L 1971 {\em Journal of Fluid Mechanics\/} {\bf 48} 41

\bibitem{sato1987}
Sato Y and Yamamoto K 1987 {\em Journal of Fluid Mechanics\/} {\bf 175} 183

\bibitem{Yeung1989}
Yeung P~K and Pope S~B 1989 {\em Journal of Fluid Mechanics\/} {\bf 207}
  531--586

\bibitem{csanady1963}
Csanady G 1963 {\em Journal of Atmospheric Science\/} {\bf 20} 201

\bibitem{Obligado2013}
Obligado M and Bourgoin M 2013 {\em New Journal of Physics\/} {\bf 15} 043019

\bibitem{Hanna1980}
Hanna S 1981 {\em Journal of Applied Meteorology\/} {\bf 20(3)} 242--249

\bibitem{Lien:jfm1998}
Lien R~C, D'Asaro E and Dairiki G 1998 {\em Journal of Fluid Mechanics\/} {\bf
  362} 177--198

\bibitem{mordant:njp2004}
Mordant N, L\'{e}v\^{e}que E and Pinton J~F 2004 {\em New Journal of Physics\/}
  {\bf 6} 116

\bibitem{Ouellette:njp2006}
Ouellette N~T, Xu H, Bourgoin M and Bodenschatz E 2006 {\em New Journal of
  Physics\/} {\bf 8} 102

\bibitem{Qureshi:prl2007}
Qureshi N~M, Bourgoin M, Baudet C, Cartellier A and Gagne Y 2007 {\em Physical
  Review Letters\/} {\bf 99} 184502

\bibitem{Brown:prl2009}
Brown R~D, Warhaft Z and Voth G~A 2009 {\em Physical Review Letters\/} {\bf
  103} 194501

\bibitem{machicoane:njp2014}
Machicoane N, Zimmermann R, Fiabane L, Bourgoin M, Pinton J~F and Volk R 2014
  {\em New Journal of Physics\/} {\bf 16} 013053

\bibitem{Zimmermann:prl2011}
Zimmermann R, Gasteuil Y, Bourgoin M, Volk R, Pumir A and Pinton J~F 2011 {\em
  Physical Review Letters\/} {\bf 106} 154501

\bibitem{bib:calzavarini2009_JFM}
Calzavarini E, Volk R, Bourgoin M, Leveque E, Pinton J~F and Toschi F 2009 {\em
  Journal of Fluid Mechanics\/} {\bf 630} 179--189

\bibitem{Volk:jfm2011}
Volk R, Calzavarini E, Leveque E and Pinton J~F 2011 {\em Journal of Fluid
  Mechanics\/} {\bf 668} 223--235

\bibitem{Monchaux:prl2006}
Monchaux R, Ravelet F, Dubrulle B, Chiffaudel A and Daviaud F 2006 {\em
  Physical Review Letters\/} {\bf 96} 124502

\bibitem{tennekes1972book}
Tennekes H and Lumley J~L 1972 {\em A first course in turbulence\/} (MIT press)

\bibitem{hinze}
Hinze J 1975 {\em Turbulence\/} (McGraw-Hill) ISBN 0070290377

\bibitem{mordant2004acc}
Mordant N, Crawford A~M and Bodenschatz E 2004 {\em Physical review letters\/}
  {\bf 93} 214501
  
\bibitem{sawford2013}Sawford, B and Pinton J~F ``A Lagrangian view of turbulent dispersion and mixing'' in Ten Chapters in Turbulence, edited by Davidson, P~A, Kaneda, Y and Sreenivasan, K~R (Cambridge University Press, Cambridge, 2013).


\end{thebibliography}

\end{document}